\documentclass{Interspeech2024}
\usepackage{algorithm}
\usepackage{algorithmic}
\usepackage{multirow}
\interspeechcameraready

\title{Complex Image-Generative Diffusion Transformer for Audio Denoising}

\name[affiliation={1}]{Junhui}{Li}
\name[affiliation={1}]{Pu}{Wang}
\name[affiliation={2}]{Jialu}{Li}
\name[affiliation={3}]{Youshan}{Zhang}

\address{
  $^1$\small Department of Mathematics, School of Science, University of Science and Technology, Liaoning, Anshan, China\\
  $^2$\small School of public policy, Cornell University, Ithaca, NY, USA. \\
  $^3$\small Department of Artificial Intelligence and Computer Science,  Yeshiva University, New York, NY, USA}
\email{Junhui\_lee@foxmail.com, 120203803006@stu.ustl.edu.cn, jl4284@cornell.edu,youshan.zhang@yu.edu}


\begin{document}

\maketitle

\begin{abstract}
    
    The audio denoising technique has captured widespread attention in the deep neural network field. Recently, the audio denoising problem has been converted into an image generation task, and deep learning-based approaches have been applied to tackle this problem. However, its performance is still limited, leaving room for further improvement. In order to enhance audio denoising performance, this paper introduces a complex image-generative diffusion transformer that captures more information from the complex Fourier domain. We explore a novel diffusion transformer by integrating the transformer with a diffusion model. Our proposed model demonstrates the scalability of the transformer and expands the receptive field of sparse attention using attention diffusion. Our work is among the first to utilize diffusion transformers to deal with the image generation task for audio denoising. Extensive experiments on two benchmark datasets demonstrate that our proposed model outperforms state-of-the-art methods.
\end{abstract}

\section{Introduction}

Audio denoising is the process of estimating better-quality audio signals from a mixture of audio by removing background noise. However, removing background noise from many different sources to produce high-quality audio still makes audio denoising challenging. Conventional audio denoising methods include Wiener filtering, spectral subtraction, minimum mean square error (MMSE) estimation~\cite{martin2005speech}, etc. In extremely low SNR and non-stationary noise environments, the performance of these approaches is known to suffer from a significant loss in performance. Deep learning techniques have resulted in multiple new audio denoising techniques to tackle challenges in the domains of speech and audio. Deep audio denoising models may estimate and remove noise from noisy data to obtain denoised audio, or they may directly create denoised audio using the regression technique~\cite{li2023deeplabv3+}.

Diffusion Denoising Probability Models (DDPM) have demonstrated great progress in generative tasks capable of generating high-quality and diverse images~\cite{peebles2022scalable}. Although these models have largely been developed in their own domains, some researchers have attempted to apply DDPM to the field of audio denoising. Zhang and Li~\cite{zhang2023complex} developed a complex image generation SwinTransformer network model to generate high-quality complex images in the Fourier domain by converting audio denoising into an image generation problem. This method has competitive performance and has outperformed previous state-of-the-art methods, such as DCU-Net~\cite{choi2018phase} and MANNER~\cite{park2022manner}. However, these methods are all based on classical UNet architectures, and their performance is still limited. Generative models include generative adversarial networks (GAN), variational autoencoders (VAEs), flow-based neural networks, and diffusion models. The GANs-based generative methods have been demonstrated to be efficient for speech enhancement~\cite{phan2020improving}. The diffusion model has been successful as a typical deep generative model. Ho et al.~\cite{ho2020denoising} first introduced a class of latent variable models motivated by nonequilibrium thermodynamics. Then, diffusion probabilistic models were used to get high-quality image synthesis results using a relatively simple architecture and training procedure. Recently, diffusion models achieved more competitive results than GANs and have achieved impressive results in various applications such as text-to-image generation~\cite{zhang2023adding}, audio synthesis~\cite{alexanderson2023listen}, and video generation~\cite{ho2022imagen}.

Most deep learning-based models for audio denoising focus on time-frequency domain (TF) methods. Due to the estimation difficulties, the majority of TF-domain approaches only accept magnitude as an input to real-valued parameter models and ignore complex-valued phases, which have an impact on performance. Although WaveNet, U-Net, or other convolutional designs serve as the basis for the majority of diffusion models, they are not scalable enough to model additional visual information. Hence, we explore a diffusion model based on a transformer with multiple inputs to capture more information and gain better complex image generation to estimate clean audio. 

To overcome the aforementioned challenges, we design a novel framework with attention diffusion on multiple input spectrograms, the complex image-generative diffusion transformer network (CIGDTN) model. We propose a complex image-generative Diffusion Transformer (CIGDT) module based on diffusion transformers (DiTs) with adaptive layer norm zero (adaLN-Zero) and sparse attention diffusion to capture more information. We also design a CIGDT block to inherit the excellent scaling properties of the transformer model and save computation costs. Furthermore, we deploy the CIGDT module to process the real and imaginary spectrograms, respectively, to make full use of the phase information in noisy audio. In addition, we apply FlashAttention-2, a novel attention algorithm with better work partitioning, to address low-occupancy or unnecessary shared memory reads and writes on the GPU~\cite{dao2023flashattention}. Overall, our contributions are threefold:

\begin{itemize}

\item  We propose a complex diffusion transformer with multiple inputs that can generate high-quality denoised audio.
\item  We present a complex image-generative diffusion transformer network (CIGDTN) model that fuses diffusion transformers with adaLN-Zero and sparse attention diffusion with FlashAttention-2 algorithm to gain better complex images.
\item Experimental results demonstrate state-of-the-art results on two benchmark datasets.
\end{itemize}

\begin{figure}
 \centering	\includegraphics[width=0.5\textwidth]{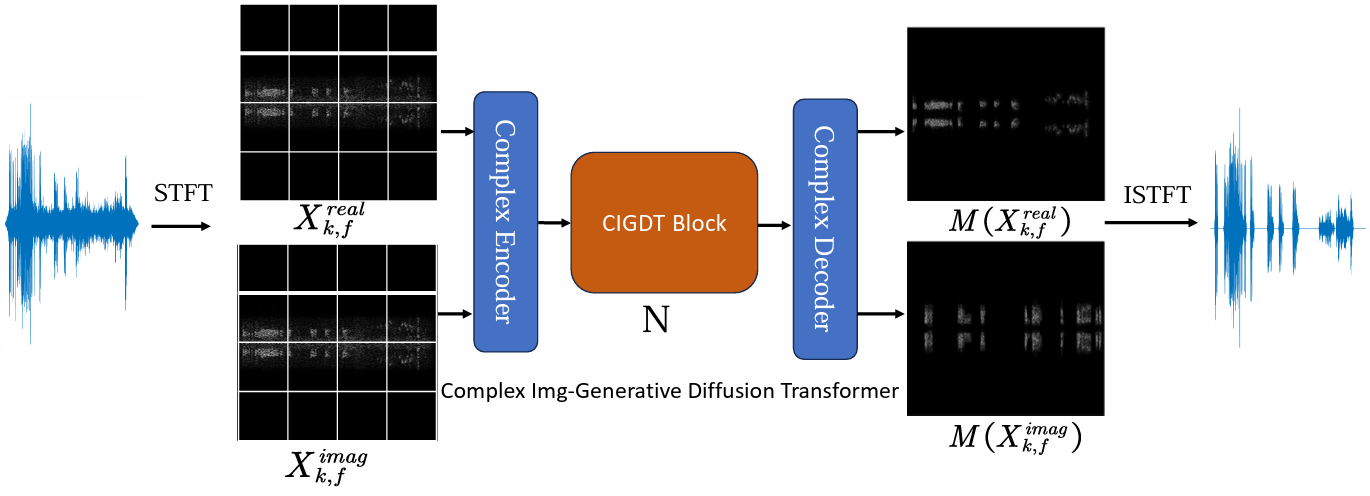}
 	%\vspace{-0.2cm}
 	\caption{A overall progress of our proposed CIGDTN model. Different modules are marked with different color blocks. The architecture of the main body fuses diffusion transformers (DiTs) architecture and sparse attention diffusion. We first apply STFT to convert audio signals $Y$ into complex images (real image $X^{real}_{k,f}$ and imaginary image $X^{imag}_{k,f}$). Then, we feed them into the model, which generates real image  $M(X^{real}_{k,f})$ and imaginary image $M(X^{imag}_{k,f})$. Finally, we reconstruct clean audio using ISTFT. }
  \label{Fig:over}
  \vspace{-0.5cm}
\end{figure}

\section{Methodology}
In audio denoising, a mixture of audio signal $y$ in the time domain can be typically expressed as a linear sum of the clean speech signal and the additive noise signal:
\begin{equation}
    y=x+\varepsilon
\end{equation}
 where $x$ and $\varepsilon$ denote clean audio and additive noise signal, respectively, a sequence of mixture signal and clean signal are defined as $Y=\{y_i\}_{i=1}^{N}$ and $X=\{x_i\}_{i=1}^{N}$, where $N$ is the total number of speech signals. Our goal is to extract a clean audio signal. Typically, each of the corresponding time-frequency $(k, f)$ audio denoising operates in the time-frequency domain:$ Y_{k,f}=X_{k,f}+\epsilon_{k,f}$, where $Y_{k,f}, X_{k,f}, \epsilon_{k,f}$ is the STFT representation of the time domain signal $y(t)$, $x(t)$, $\varepsilon(t)$ and $k$, $f$ are the time frame index and frequency bins index. 

\section{CIGDTN Model}
To reduce noise and recover the speech signal, this study only concentrates on the denoising tasks in the Fourier domain. Additionally, we employ a complicated feature encoder trained end-to-end to enhance the information of various image bands rather than just using STFT features as input. We also restore the features to the time-frequency domain using a complex feature decoder. To accomplish this task, we developed a complex image-generative diffusion transformer network (CIGDTN) model to handle complex image inputs. The model architecture can be found in Fig.~\ref{Fig:over}, which mainly consists of three main parts: a complex encoder, a CIGDT module, and a decoder. \\
\begin{figure*}
 \centering	\includegraphics[width=0.9\textwidth]{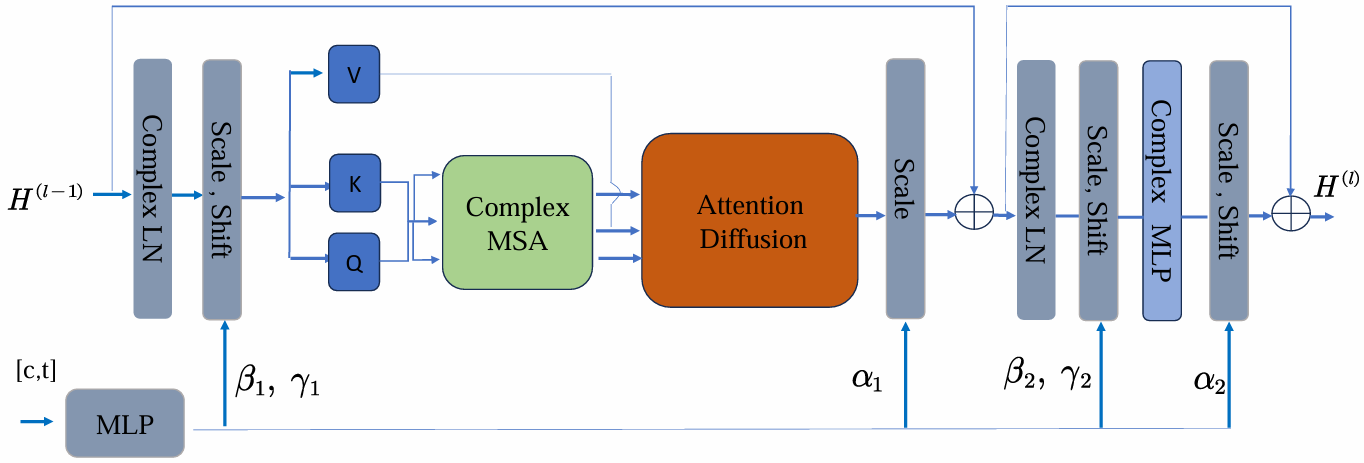}
 	%\vspace{-0.2cm}
 	\caption{CIGDT block architecture. Each block is composed of a complex Multi-Head Self-Attention (MSA) module and sparse attention diffusion, and a complex feed-forward layer including a complex LayerNorm layer, a two-fully connected layers complex MLP (CMLP). A scale and shift layer is located behind the LayerNorm layer.}
  \label{Fig:CIGDT}
  \vspace{-0.5cm}
\end{figure*}
To make full use of the phase information in noisy audio, our model takes all the real and imaginary spectrograms as inputs. Given an input batch of tensor $T^r, T^i \in \mathbb{C}^{N\times C\times H\times W}$, where $N$ and $C$ is the number of samples in the batch and the channel size respectively; $H$ and $W$ represent the height and width of the image.

% \subsection{Complex separation operations.} A complex-valued tensor is defined as $T=A+iB$ with real-valued matrices $A$ and $B$, which represent the real part $T^r$ and imaginary part $T^i$ of a complex tensor, respectively. The basic deep learning operations $O(*)$ in the complex field can be defined as:
% \begin{equation}\label{comeq}
%     \mathcal{C}omplexO(T)= O(A)+iO(B),
% \end{equation}
% % \begin{equation}
% % \begin{aligned}
% %     W\times h=(A\times x-B\times y)+i(B\times x+A\times y) 
% % \end{aligned} 
% % \end{equation}

% where $\mathbb(C)$ denote a complex function, $i$ represents the imaginary unit, and common deep learning operations $O(*)$ such as Conv2d, MaxPool2d, BatchNorm2d, ReLU, Dropout, LayerNorm, Linear, etc. are also adapted to the complex domain by applying Eq.~\ref{comeq}. we can get the complex version of these deep learning operations. This shows that any complex function can be rewritten as two separate real functions.

\subsection{Complex Encoder}
Complex-valued input audio is translated into a complex-valued representation by the complex encoder. The complex neural network has shown promising performance due to its effectiveness in processing complex-valued spectrograms.  To start with, we transform the raw real-valued audio $(y_1,...,y_n)\in \mathbb{R}^{d\times n}$ into complex-valued tensor $\mathbf{T}_1,...,\mathbf{T}_n$ by STFT, an operation which decomposes a finite time sequence into a finite frequency sequence to generate a complex tensor representation for audio images. Given a complex-valued tensor $T$, we use our model to convert the input into a sequence of patches $\mathbf{T}=[T_1...T_N]\in \mathbb{R}^{N\times P^2 \times C}$ tokens where $(P,P)$ is the patch size, $N=HW/P^2$ is the number of patches. The number of tokens $T$ created by patchify is determined by the patch size hyperparameter $p$. Following patchify, we apply specific position embeddings $Pos=[Pos_1,..., Pos_N]\in \mathbb{R}^{N\times D}$ to all input tokens to retain positional information as follows:
\begin{equation}
    \mathcal{Z}=[T_1{E};T_2{E},\cdots;T_N{E}]+E_{Pos}
\end{equation}
where $E\in \mathbb{R}^{(P^2\cdot C)\times D}$ is the patch embedding projection, and $E_{Pos} \in \mathbb{R}^{N\times D}$ denotes the position embedding.

\subsection{CIGDT Block}
Following the encoder, the input tokens are processed by a sequence of transformer blocks. The encoder consists of four CIGDT blocks. The key design feature of the CIGD transformer is to design a DiT with adaLN-Zero and sparse attention with Attention Diffusion.
We replace standard layer norm layers in transformer blocks with adaptive layer norm (adaLN). Rather than directly learning dimensionwise scale and shift parameters $\gamma$ and $\beta$, we regress them from the sum of the embedding vectors of $t$ and $c$. We also use a similar initialization strategy as diffusion U-Net models, zero-initializing the final convolutional layer in each block prior to any residual connections. To further broaden the receptive field of sparse attention, attention diffusion is used. This technique computes multi-hop token correlations based on all pathways between corresponding disconnected tokens in addition to the attention between surrounding tokens. An attention layer, a feed-forward layer with a LayerNorm layer, a two-layer MLP, and GELU nonlinearity are components of every CIGDT block. A scale and shift layer is located behind the LayerNorm layer. Additionally, we add regression dimensionwise scaling parameters $\alpha$, which are applied immediately prior to any residual connections within the CIGDT block.

With such a transformer architecture and adaLN scheme, consecutive CIGDT blocks can be formulated as:
\begin{equation}
   \begin{aligned} 
   &\hat{\mathcal{H}}^l=\mathbb{C}MSA(\gamma \mathbb{C}LN(\mathcal{H}^{l-1})+\beta)+\mathcal{H}^{l-1}\\
    &\mathcal{H}^l=\mathbb{C}MLP(\gamma \mathbb{C}LN(\hat{\alpha \mathcal{H}}^l)+\beta)+\hat{\mathcal{H}}^l
   \end{aligned} 
\end{equation}
 where $\hat{\mathcal{H}}^l$ and $\mathcal{H}^l$ denote the output features of the MSA module and the MLP module for block $L_i$, respectively. $\gamma$, and $\beta$ denote demension-wise scale and shift parameters for LayerNorm layer. $\alpha$ denotes dimensionwise scaling parameters that are applied immediately prior to any residual connections.
 
\textbf{Sparse Attention Diffusion.} By combining token correlations that are multiple hops away, sparse attention diffusion was developed. Especially, sparse patterns consider a combination of local window attention, global attention, and random attention to capture token interactions without quadratic complexity dependency on the sequence length. The attention matrix $A$ is first used to characterize the interaction strength between neighboring nodes on the graph $G$, i.e.,
\begin{equation}
A_{i,j}=\frac{exp(Q_{i}K_{j})/\sqrt{d}}{\sum_{j\in Ne(i)}exp(Q_{i}K_{j})/\sqrt{d}}.
\end{equation}

Sparse attention diffusion utilizes the attention diffusion process to calculate the multi-hop token relationships on the attention graph based on attention weights on edges. The entries of the graph diffusion matrix $\mathcal{A}$ are calculated to get the multi-hop attention scores:
\begin{equation}
    \mathcal{A}=\sum^{\infty}_{s=0}\delta_{k}A^k,
\end{equation}
where $A$ is the calculated sparse attention matrix, and the weighting coefficient $\delta_k$ satisfies $\sum^{\infty}_{s=0}\delta_{k}=1, \delta_k\in[0,1]$. The sparse attention pattern's original receptive fields will gradually enlarge as $k$ increases. All paths between tokens $i$ and $j$ are included in the resulting attention score $A_{i,j}$ and weighted by the coefficient $\delta_k$. Next, We multiply the diffusion attention matrix $A$ by each value vector $V$ as Eq.\eqref{Eq:att}.

\begin{equation} \label{Eq:att}
Attention(Q,K,V)=softmax(\frac{QK^T}{\sqrt{d_k}})V=AV
\end{equation}
where $d_k$ denotes the dimension of $K$.

Even when the sparsity is taken into account, computing the power of attention matrices can be unavoidably expensive for long sequences. To efficiently combine the diffusion mechanism with transformers, we implement the graph diffusion process as the personalized pagerank (PPR) by specifying $\delta_k=\alpha(1-\alpha)^k$ with teleport probability $\alpha$. The resulting diffusion matrix $\mathcal{A}=\sum^{\infty}_{k=0}\alpha(1-\alpha)^{k}A^{k}$ is the power expansion of the solution to the recursive equation $\mathcal{A}=\alpha I+(1-\alpha)\mathcal{A}A$. Each power diffusion step is calculated as
\begin{equation}
\begin{aligned}
    & Z_0=V=XW_v, \\
    & Z_{k+1}=(1-\alpha)AZ_k+\alpha V,
\end{aligned}
\end{equation}
for $0\leq k<K$. $Z_k$ is the output of the attention diffusion process and will converge to the real output $\mathcal{A}V $ as $K\rightarrow \infty$.  

\subsection{Decoder} 
After the final CIGDT block, we need to decode our sequence of generated audio image tokens into an output-denoised image prediction. Both of these output images have shapes that are equal to the original spatial input. We apply the final layer norm (adaptive if using adaLN) and linearly decode each token into a $p\times p\times 2C$ tensor, where $C = 1$ is the number of channels in the spatial input to CIGDT module. The term ``channel`` is often used in the field of image processing, and a channel number of 1 usually indicates a grayscale image. Finally, we rearrange the decoded tokens into their original spatial layout to get the predicted denoised image. After getting the output from the decoder layers in the CIGDTN model, we could apply ISTFT to get the reconstructed audio as $\hat{Y}$.
The overall training algorithm is shown in Alg.~\ref{alg:AGTD}.
\begin{algorithm}[ht]
    \caption{CIGDTN: Complex Image-Generative Diffusion Transformer Network for Denoising. Batch of audio input: $B(U)=\{U^1,...,U^{n_B}\}$, where ${n_B}$ is the total number of batch. $I$ is the number of iterations }
    \label{alg:AGTD}
 \begin{algorithmic}[1]
    \STATE {\bfseries Input:} Mixture audio signals $Y=\{y_i\}_{i=1}^{NN}$ and raw audio input $X=\{x_i\}_{i=1}^{NN}$, where $NN$ is the total sample number of audios.
    \STATE {\bfseries Output:} Denoised audio signals: $\hat{X}$
    \STATE Converting to mixture audio image $X_{k,f}=\{I_i\}^{NN}_{i=1}$ and raw audio image $Y_{k,f}=\{Y_i\}^{NN}_{i=1}$ through STFT.
    \FOR{$iter =1$ {\bfseries to} $I$}
        \FOR{$j =1$ {\bfseries to} $n_B$}
          \STATE Derive batch-wise data: $X_{k,f}^j$ and $Y_{k,f}^j$ 
    \STATE Obtain the image loss by Calculating Eq.~\eqref{eq:imgloss}   
    \STATE Restore complex images to audio signals using ISTFT and Obtain audio loss by Calculating.\eqref{eq:SDRloss}
    \STATE Optimize our audio generation model CIGDT network by Calculating Eq.~\eqref{eq:loss}
         \ENDFOR
    \ENDFOR
    \STATE Using ISTFT to output denoised audio signals
 \end{algorithmic}
\end{algorithm}

\subsection{Objective Function}
In this study, our model processes real- and imaginary-image streams to extract more audio features. Then, the estimated output is reconstructed by ISTFT. Therefore, our loss function consists of image loss and SDR loss to fully utilize different feature information. The image loss is based on the energy-conserving loss function proposed, which simultaneously considers clean audio complex images and noisy audio complex images. We first apply $L_1$ loss to minimize the difference between the generated images and the ground truth image $Loss_{im,L_1}=| y-\hat{y} |_1$ Eventually, the image loss consists of three parts and is defined as follows:
\begin{equation}\label{eq:imgloss}
    Loss^{total}_{L_1,im}=Loss^{real}_{im,L_1}+Loss^{imag}_{im,L_1}+|\varepsilon-\hat{\varepsilon} |_1,
\end{equation}
where $y$ and $\hat{y}$ are the samples of the clean audio complex images and the enhanced audio complex images, respectively. $\varepsilon$ represents the additive noise signal given a mixture of the audio signal and $\hat{\varepsilon}=x-\hat{y}$ represents the estimated noise and $|\cdot|$ denotes the $L_1$ norm.

For the reconstructed audio signal, we also first apply $L_1$ loss to minimize the difference between the reconstructed audio and the ground truth audio as follows: 
\begin{equation}\label{eq:SDRloss}
    Loss_{R,L_1}=|Y-\hat{Y} |_1
\end{equation}

To properly balance the contribution of these two loss terms and to address the scale insensitivity problem, we weigh each term proportionally to the energy of each utterance. The final form of the loss function is as follows:
\begin{equation}
\begin{aligned}\label{eq:loss}
    Loss^{total}=\alpha Loss^{total}_{L_1,im}+(1-\alpha) Loss_{R,L_1}
\end{aligned}
\end{equation}

\section{Experiment}

We evaluated the proposed CIGDT with two audio datasets, VoiceBank+DEMAND and Birdnoisesound dataset. The model was trained for 100 iterations on a single NVIDIA 3060 GPU. We train all models with AdamW. We use a constant learning rate of $1\times 10^{-4}$, no weight decay, and a batch size of 8. In order to address low-occupancy or unnecessary shared memory reads and writes on the GPU, we employ the FlashAttention-2 algorithm. In order to convert audio signals into audio images, we used the STFT and a 500-point Hamming window function with a Fourier transform of $nfft=513$. Each audio's length can be different. Therefore, we set the distance between neighboring sliding window frames to be $hop_{length}=int(length(x_t)/256)$, where $length(x_t)$ is the length of each audio. The input image dimensions are then resized as $[256 \times 256 \times 1]$.
\subsection{Datasets}
\textbf{VoiceBank+DEMAND} is a synthetic dataset created by mixing clean speech and noise~\cite{valentini2016speech}. The training set contains 11572 utterances (9.4h), and the test set contains 824 utterances (0.6h). The lengths of utterances range from 1.1s to 15.1s, with an average of 2.9s. \\
\textbf{BirdSoundsDenoising} was randomly split into the training set (10000 samples), validation set (1400 samples), and test set (2720 samples)~\cite{zhang2023birdsoundsdenoising}. Unlike many audio-denoising datasets, which have manually added artificial noise, these datasets contain many natural noises, including wind, waterfall, rain, etc.

\begin{table}[!ht]
  
  \centering
  %\vspace{-0.1cm}
\setlength{\tabcolsep}{+0.5mm}{
  \begin{tabular}{lclllll}
    \toprule
    Methods  & Domain & PESQ & STOI & CSIG & CBAK & COVL  \\
    \hline
% Wave-U-Net~\cite{macartney2018improved} &T &2.62 &- &3.91 &3.35 &3.27 \\

% MMSE-GAN~\cite{soni2018time} &TF &2.53 &0.93 &3.80& 3.12& 3.14 \\
PGGAN~\cite{li2022perception} &T &2.81 &0.944& 3.99& 3.59 &3.36 \\
DCCRGAN~\cite{huang2022dccrgan} &TF &2.82 &0.949 &4.01& 3.48 &3.40 \\
%S-DCCRN~\cite{lv2022s} &TF &2.84 &0.940 &4.03 &2.97 &3.43 \\
% HiFi-GAN~\cite{su2020hifi} &T &2.94 &- &4.07 &3.07& 3.49 \\ 
%TSTNN~\cite{wang2021tstnn} & T & 2.96 & 0.950 & 4.33 & 3.53 & 3.67\\
%PHASEN~\cite{yin2020phasen} &TF &2.99 &$-$ &4.18& 3.45& 3.50 \\
%DEMUCS~\cite{defossez2020real} & T & 3.07 &0.95& 4.31 & 3.40 & 3.63 \\
SE-Conformer~\cite{kim2021se} & T & 3.13 &0.95& 4.45 & 3.55 & 3.82 \\
MetricGAN+~\cite{fu2021metricgan+} &TF &3.15 &0.927& 4.14& 3.12 &3.52\\
MANNER~\cite{park2022manner} & T & 3.21 & 0.950 & 4.53 & 3.65 & 3.91\\
CMGAN~\cite{cao2022cmgan} & T & 3.41 &0.96& 4.63 & 3.94 & 4.12 \\
DPATD~\cite{li2023dpatd} & T & 3.55 &0.97 & 4.78 & 3.96 & 4.22 \\
CIGSN~\cite{cao2022cmgan} & TF & 3.41 &0.954& 4.78 & 3.82& 4.22 \\
\hline
\textbf{CIGDTN} & TF & \textbf{3.55} & \textbf{0.964} & \textbf{4.82} & \textbf{3.96} & \textbf{4.24} \\
% \bottomrule
\hline
  \end{tabular}}
  \caption{Comparison results on the VoiceBank-DEMAND dataset. ``$-$" means not applicable.}
  \label{tab:voice}
  \vspace{-0.9cm}
\end{table}

\begin{table}[!ht]
\small
\centering
%\captionsetup{font=small}
% \vspace{-0.2cm}
\setlength{\tabcolsep}{+0.5mm}{
\begin{tabular}{lllll|lllllllll}
\hline \label{tab:md}
 \multirow{2}{*}{Networks}
 &  \multicolumn{4}{c}{Validation} & \multicolumn{4}{c}{Test} \\
 \cmidrule{2-9}
& $F1$ & $IoU$  & $Dice$ & $SDR$ & $F1$ & $IoU$  & $Dice$ & $SDR$ \\
\hline
U$^2$-Net~\cite{qin2020u2}  &60.8 &45.2 &60.6 &7.85 & 60.2  &44.8 &59.9 & 7.70\\
MTU-NeT~\cite{wang2022mixed}  &69.1 &56.5 &69.0  &8.17 & 68.3  &55.7 & 68.3 &7.96  \\
Segmenter~\cite{strudel2021segmenter} & 72.6  & 59.6 & 72.5 & 9.24 & 70.8 & 57.7 & 70.7 & 8.52   \\
%U-Net~\cite{ronneberger2015u}  &75.7 &64.3 &75.7 & 9.44 &74.4 &62.9 &74.4 & 8.92    \\
% DeepLabV3  & 83.4 & 83.4 & 75.9  &  & 83.0 & 83.0 & 75.4 & 82.6 \\
SegNet~\cite{badrinarayanan2017segnet}  &77.5 &66.9 &77.5 & 9.55&76.1 &65.3 &76.2 & 9.43 \\
DVAD~\cite{zhang2023birdsoundsdenoising} & 82.6  & 73.5 & 82.6 & 10.33  & 81.6 & 72.3 & 81.6 & 9.96 \\
R-CED~\cite{park2016fully} & $-$ & $-$ & $-$ &2.38     &$-$ &$-$&$-$ & 1.93  \\
Noise2Noise~\cite{kashyap2021speech}  & $-$ & $-$ & $-$ & 2.40&$-$ &$-$&$-$ &1.96\\
TS-U-Net~\cite{moliner2022two}  & $-$ & $-$ & $-$ & 2.48&$-$ &$-$&$-$ &1.98\\
DCHT~\cite{li2023dcht} & $-$ & $-$ & $-$ &10.49    &$-$ &$-$&$-$ & 10.43  \\
CIGSN~\cite{zhang2023complex} & $-$ & $-$ & $-$ &10.69    &$-$ &$-$&$-$ & 10.15  \\
\hline
\textbf{CIGDTN} & $-$  & $-$ & $-$ &  \textbf{10.65}  & $-$ & $-$ & $-$ & \textbf{10.25} \\
\hline
\end{tabular}}
\caption{Results comparisons of different methods on Birdsoundsdenoising dataset. ($F1, IoU$, and $Dice$ scores are multiplied by 100. ``$-$" means not applicable.) }
\label{sdr}
\vspace{-0.5cm}
\end{table}

\subsection{Evaluation Metrics}\label{sec:me}
For evaluation on the Birdsoundsdenoising dataset, we use signal-to-distortion ratio (SDR) to evaluate different models. 
%
% The model is better when these metrics are greater.
% \begin{equation}\label{eq:sdr}
%     SDR=10log_{10}\frac{||u||^2}{||\widetilde{u}-u||^2}
% \end{equation}
% where $u$ is the ground truth labeled masks of audio image and $\widetilde{u}$ is the predictions of segmentation.
%
We assess the proposed audio denoising model on the VoiceBank+DEMAND dataset using a variety of objective metrics: perceptual evaluation of speech quality (PESQ, higher is better) with a score range from -0.5 to 4.5; short-time objective intelligibility (STOI, higher is better) with a score range from 0 to 1. We also adopt subject mean opinion scores (MOSs; higher is better), such as CSIG for evaluating signal distortion, CBAK for evaluating noise distortion, and COVL for evaluating overall quality. 

\subsection{Result}
Table~\ref{tab:voice} shows the comparison results of our proposed model and SOTA baselines on the VoiceBank+DEMAND dataset. As we can see, CIGDTN surpasses most waveform-based approaches currently in use in all five metrics and performs as well as other methods with large model configurations while employing fewer parameters. For the BirdSoundsDenoising dataset, we report the performance of our CIGDTN model and ten state-of-the-art baselines. The results are shown in Table~\ref{sdr}, where the bold text indicates the best outcomes. The results demonstrate that our model outperforms other state-of-the-art methods in terms of SDR. Results of F1, IoU, and Dice are omitted since these metrics are used for the audio image segmentation task~\cite{zhang2023birdsoundsdenoising}. As a consequence, these benchmarks confirm that our method for audio denoising is effective, and our model enhances the audio-denoising performance of both VoiceBank+DEMAND and BirdSoundDenoising datasets.

\section{Conclusion}
In this paper, we present a complex image-generative diffusion transformer network (CIGDTN) model for audio denoising. CIGDTN explores a new class of diffusion models based on transformer architecture with multiple inputs to achieve better complex image generation and audio denoising. In a CIGDT block, diffusion transformers were improved with sparse attention. We also modified the transformer model using the FlashAttention-2 algorithm, which can compute attention with a great deal fewer memory accesses. Extensive experiments on two benchmark datasets demonstrate the superiority of the proposed CIGDTN architecture in audio-denoising tasks.

\bibliographystyle{IEEEtran}
\bibliography{mybib}

\end{document}